\documentclass[a4paper,11pt]{article}
\usepackage{pos}
\usepackage{bm}
\usepackage{placeins}

\title{High-$x$ quark density and their impact on $Z'$-boson dilepton searches} \ShortTitle{$Z'$-boson searches and high-$x$ PDFs}

\author*[a]{Francesco Giuli}

\affiliation[a]{CERN, CH-1211 Geneva, Switzerland}

\emailAdd{francesco.giuli@cern.ch}

\abstract{In this proceeding, we study the influence of theoretical systematic uncertainties due to the quark density on LHC experimental searches for $Z'$-bosons. Using an approach originally proposed in the context of the ABMP16 PDF set for the high-$x$ behaviour of the quark density, we presents results on differential cross section and Forward-Backward asymmetry observables commonly used to study $Z'$ signals in dilepton channels.}

\FullConference{The European Physical Society Conference on High Energy Physics (EPS-HEP2023)\\
 21-25 August 2023\\
Hamburg, Germany\\}

\begin{document}
\maketitle

\section{Introduction}
\begin{figure}[t!]
\begin{center}
\includegraphics[width=0.40\textwidth]{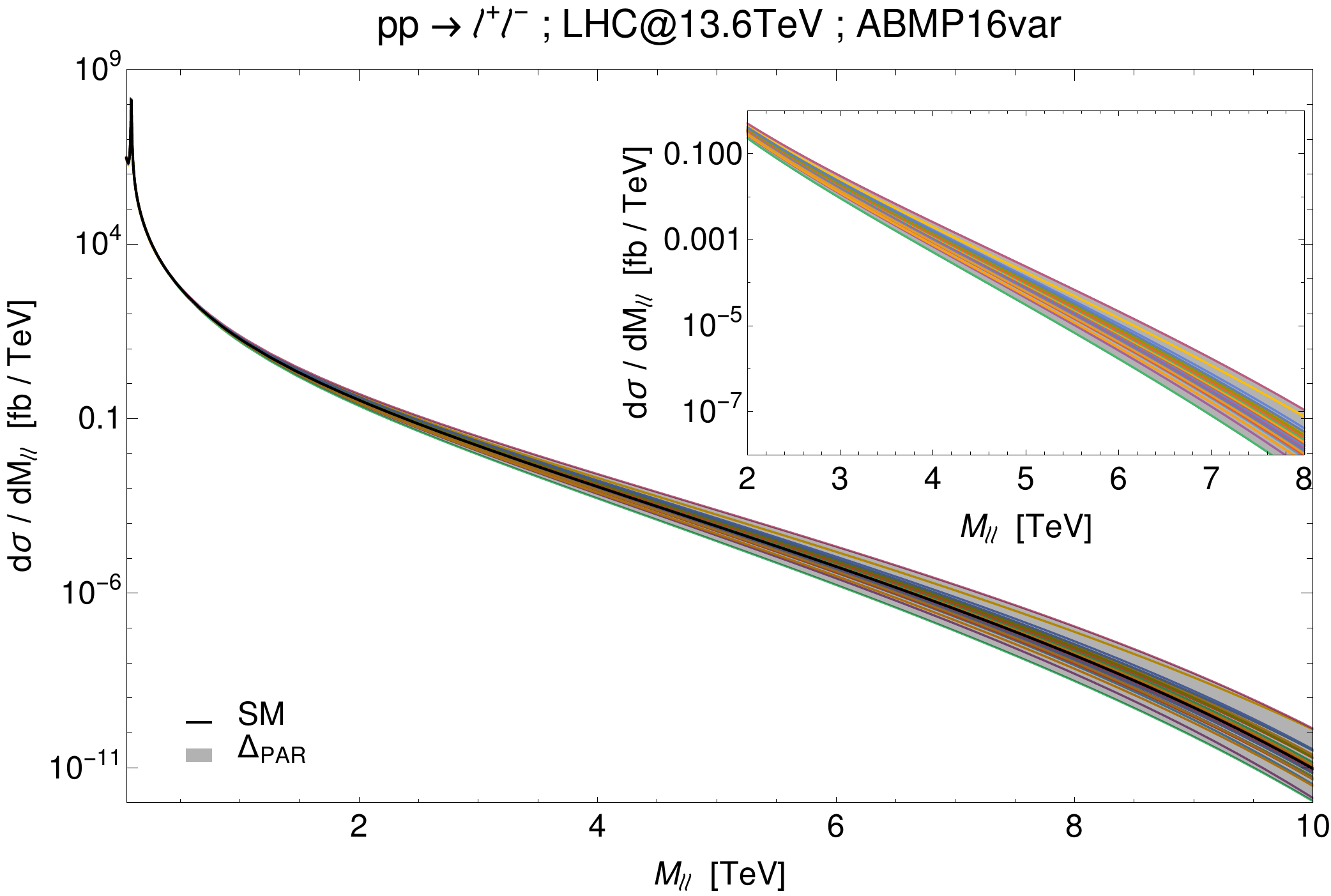}
\includegraphics[width=0.40\textwidth]{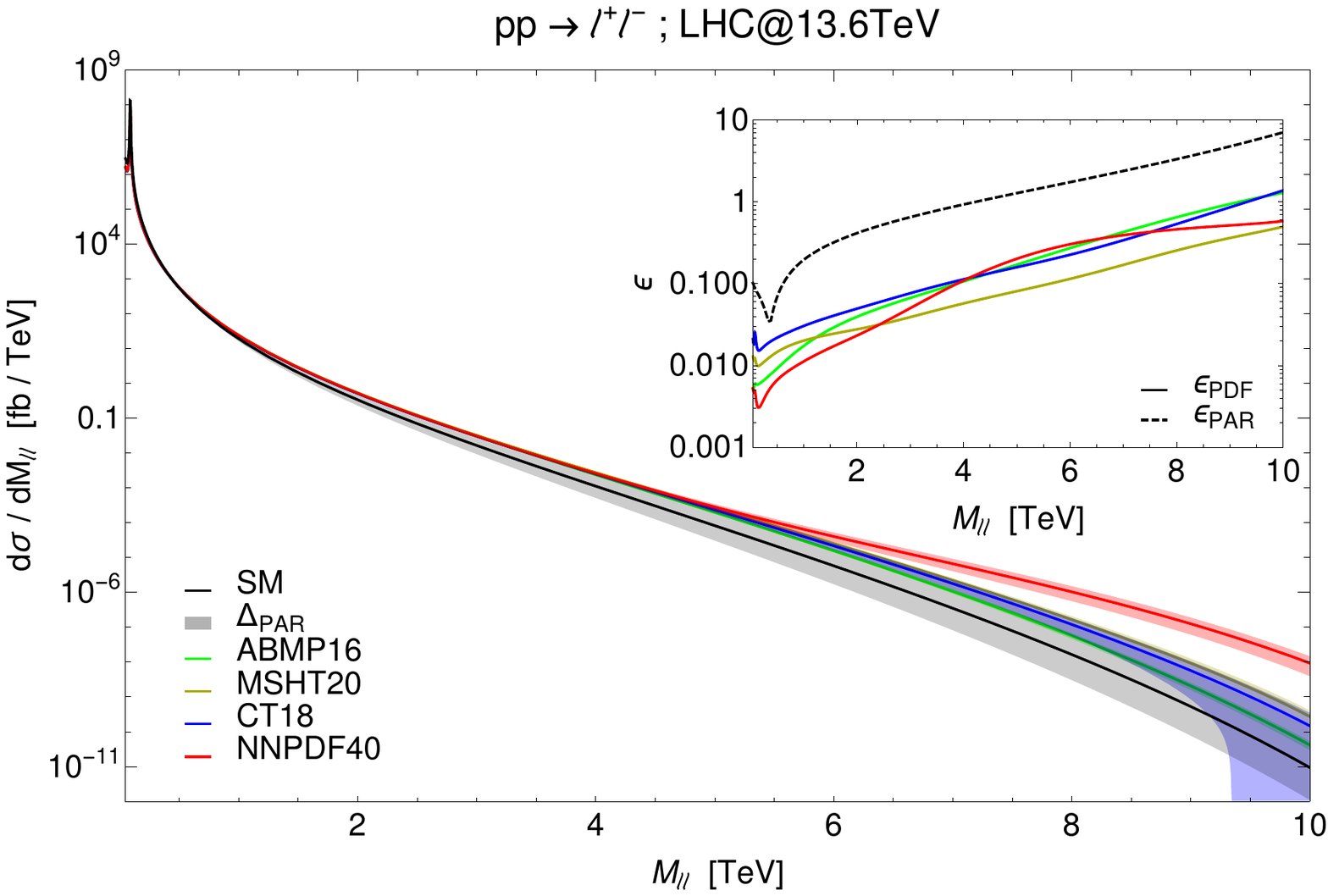}
\caption{Differential cross section in invariant mass of the dilepton final state at the LHC with $\sqrt{s}$ = 13.6 TeV:  
 (left) results for all ABMP16var members, with the inset plot 
 focusing on the few TeV invariant mass region; (right) ABMP16var result 
 compared with results from standard PDF sets,  
 with the inset plot showing the relative sizes of parameterisation and ``standard" PDF uncertainties.}
\label{fig:invmass}
\end{center}
\end{figure}
The Parton Density Functions (PDFs), which describe the Quantum Chromo-Dynamics (QCD) evolution of the initial partonic states entering the collision,  represent one of the main sources of systematic uncertainties, affecting the potential of experimental searches for discovering or setting exclusion bounds on new $Z^\prime$ bosons. The crucial role of the PDFs can be illustrated, e.g., by the analyses~\cite{Fiaschi:2021sin,Fiaschi:2021okg,Accomando:2019vqt,Accomando:2018nig,Accomando:2017scx} of Drell-Yan (DY) processes, i.e., dilepton channels in hadronic collisions. \\
Further aspects of the PDF systematics in the $Z^\prime$-boson dilepton search region are investigated in this proceeding. We concentrate on the effect of the $x \to 1$ behaviour of the quark densities, as already pointed out in Ref.~\cite{Alekhin:2017kpj}. The very high mass tails of physical distributions are influenced by the quark density at high $x$ and low mass scales through QCD evolution. This sensitivity can be recast as a parameterisation uncertainty $(1-x)^b$ in the nearly-vanishing quark density as $x \to 1$~\cite{Courtoy:2020fex}.
This proceeding reviews the results presented in Ref.~\cite{Fiaschi:2022wgl}, where we followed the approach proposed in~\cite{Alekhin:2017kpj} to treat the $x \to 1$ quark density and analysed its impact on the multi-TeV mass region relevant for BSM $Z^\prime$ searches. 

\section{Quark densities at large-\texorpdfstring{$\boldsymbol{x}$}{x}}
\begin{figure}[t!]
\begin{center}
\includegraphics[width=0.40\textwidth]{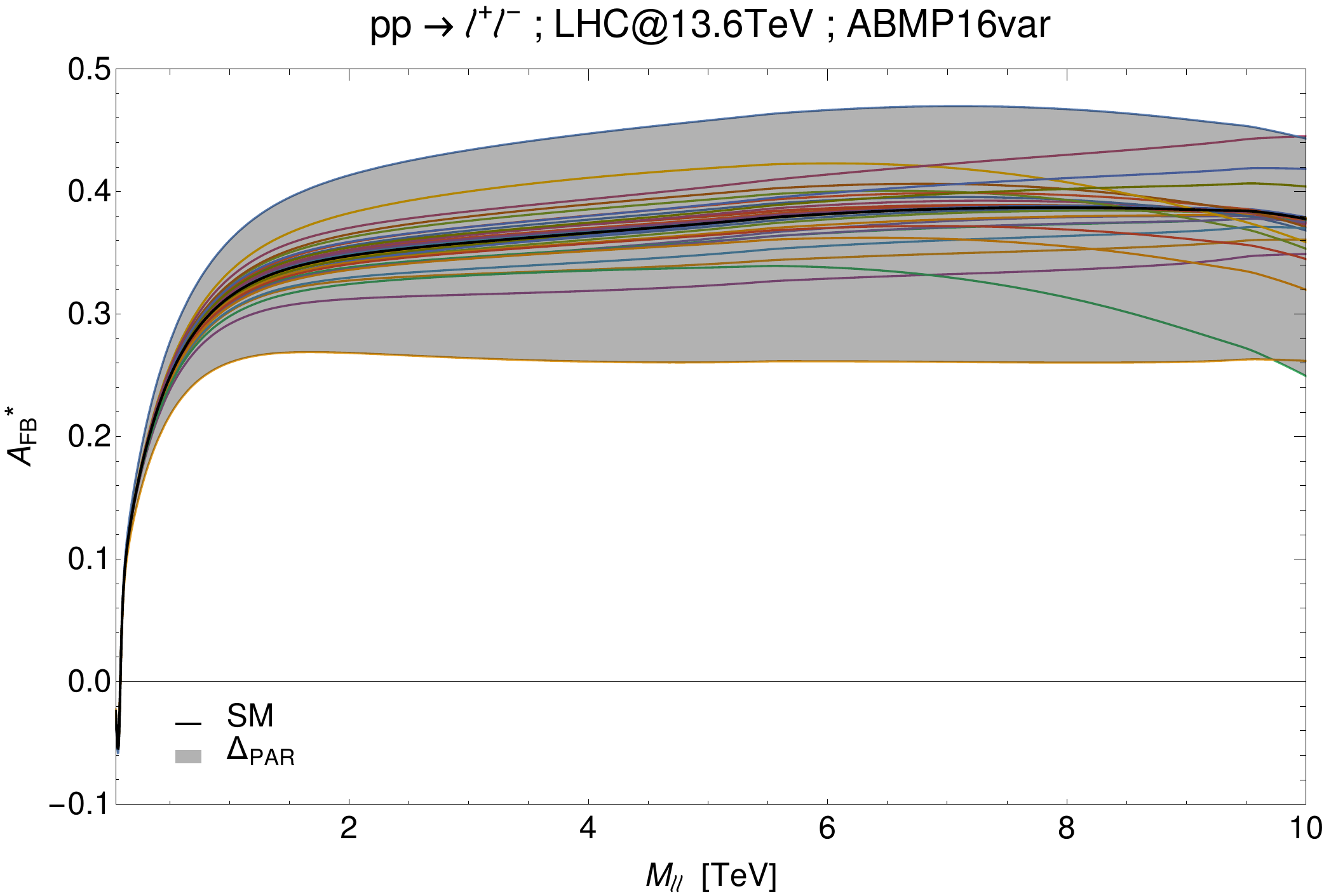}
\includegraphics[width=0.40\textwidth]{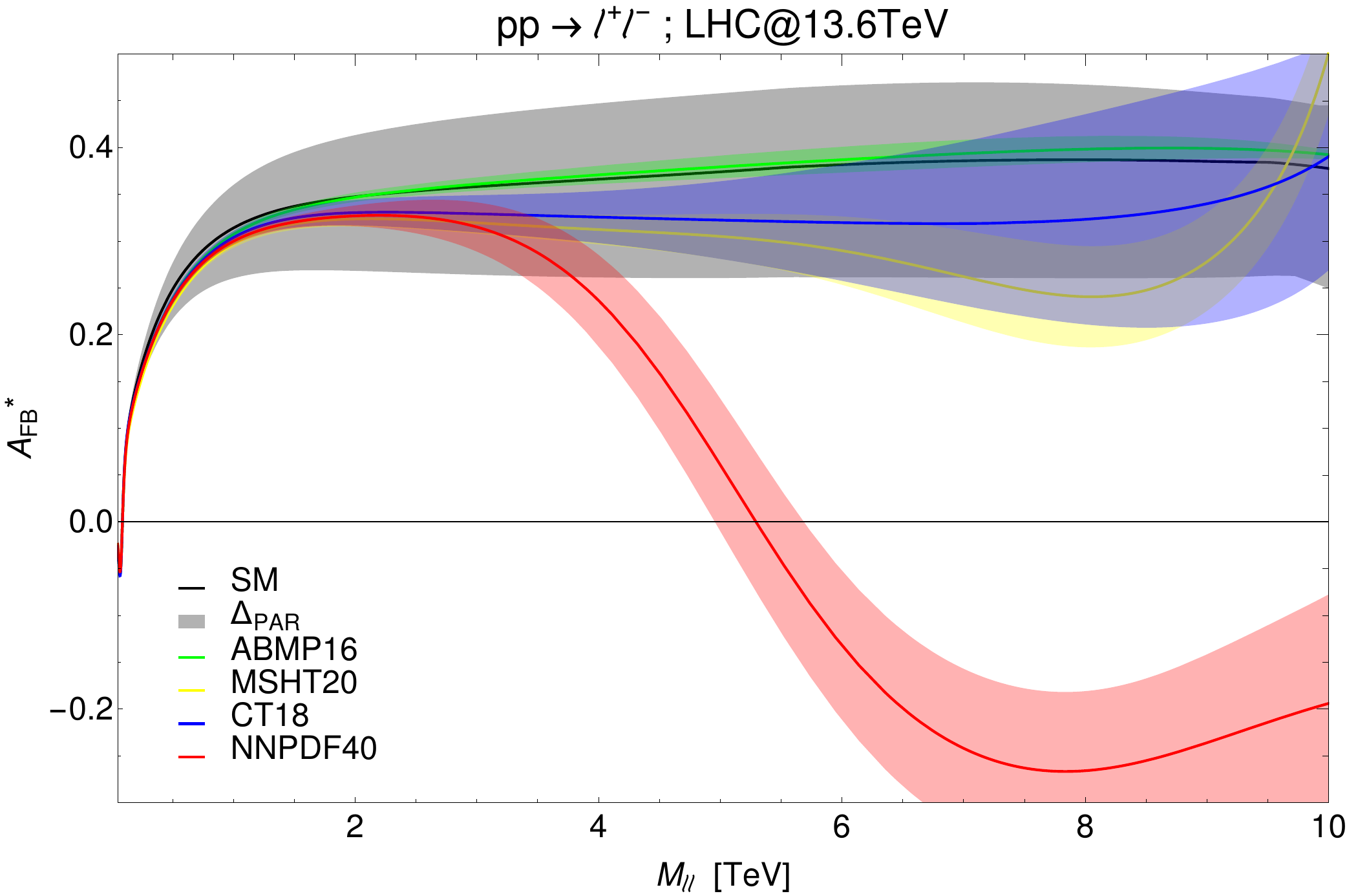}
\caption{Asymmetry $A_{\rm{FB}}^*$ in 
invariant mass of the dilepton final state at the LHC with $\sqrt{s}$ = 13.6 TeV:  (left) results for all ABMP16var members; (right)  ABMP16var result compared with results from standard PDF sets.}
\label{fig:afb-SM}
\end{center}
\end{figure}
To study the impact of high-$x$ quark densities on dilepton observables, we construct a PDF ensemble as follows, by using the \texttt{xFitter} framework~\cite{Alekhin:2014irh,xFitter:2022zjb}.
The ``central'' value of the ensemble is obtained using the results of the fit contained in Tab.VII of Ref.~\cite{Alekhin:2017kpj}.
We then construct 24 ensemble members by varying the exponent of the $(1 - x)$ term of the parameterisation of $u$ and $d$ quarks and anti-quarks by $\pm$ 0.3, 0.5, 1.0.
We obtain  predictions for dilepton observables using 22 members of the set,  excluding the members with  variations by $+1$ for $u$ and by $-1$ for $d$ as these do not yield a sufficiently fast fall-off in the ratio $d / u$ for $x \to 1$. We further include predictions obtained from the following additional members:
\begin{itemize}
\item All quark and anti-quark distributions varied simultaneously by $\pm$ 0.3, 0.5, 1 (i.e., 6 additional variations). We name this Variation \#1.
\item Quark distributions varied by $\pm$ 1  and anti-quark distributions 
varied by $\mp$ 1  simultaneously (i.e., 2 additional variations).
We name this Variation \#2.
\end{itemize}
All these combinations ensure a vanishing $d/u$ ratio as $x \to 1$.
In the following the predictions obtained from the PDF ensemble 
described above will be referred to as ``ABMP16var'', and the envelope from the 30 different variations will be used as an estimate of the parameterisation uncertainty.

\FloatBarrier\section{SM results}
We then computed the dilepton invariant mass distribution of the 
differential cross section $ d \sigma / d M_{\ell \ell} $  and 
forward-backward asymmetry $A_{\rm{FB}}^*$.  Fig.~\ref{fig:invmass} shows the results for $ d \sigma / d M_{\ell \ell} $, while Fig.~\ref{fig:afb-SM} for $A_{\rm{FB}}^*$. 
The results for all the variations in the ABMP16var ensemble, and the parameterisation uncertainty $\Delta_{\rm{PAR}}$ from their envelope., are shown in the left panels of these figures.  For  $ d \sigma / d M_{\ell \ell} $ the  largest variations are obtained from  Variation \#1, while for 
$A_{\rm{FB}}^*$ the largest variations are obtained from  Variation \#2.\\
The ABMP16var results are compared with the results from 
the original ABMP16 set~\cite{Alekhin:2017kpj} and other commonly used sets  CT18~\cite{Hou:2019efy}, MSHT20~\cite{Bailey:2020ooq} and 
NNPDF4.0~\cite{NNPDF:2021njg} in the right panels of Figs.~\ref{fig:invmass} and~\ref{fig:afb-SM}. It can be seen that the  relative sizes $\epsilon = \Delta\sigma / \sigma$ of the parameterisation  uncertainties ($\epsilon_{\rm{PAR}}$) is roughly one order of magnitude larger than PDF uncertainties of  each set ($\epsilon_{\rm{PDF}}$). \\
All the predictions for the SM differential cross section  agree for  all PDF sets  within uncertainties, except NNPDF4.0 which departs from the others for $ M_{\ell \ell} \gtrsim  5$ TeV.  This peculiar behaviour has been investigated in the recent work of Ref.~\cite{Ball:2022qtp}. 

\section{BSM searches}
\begin{figure}[t!]
\begin{center}
\includegraphics[width=0.40\textwidth]{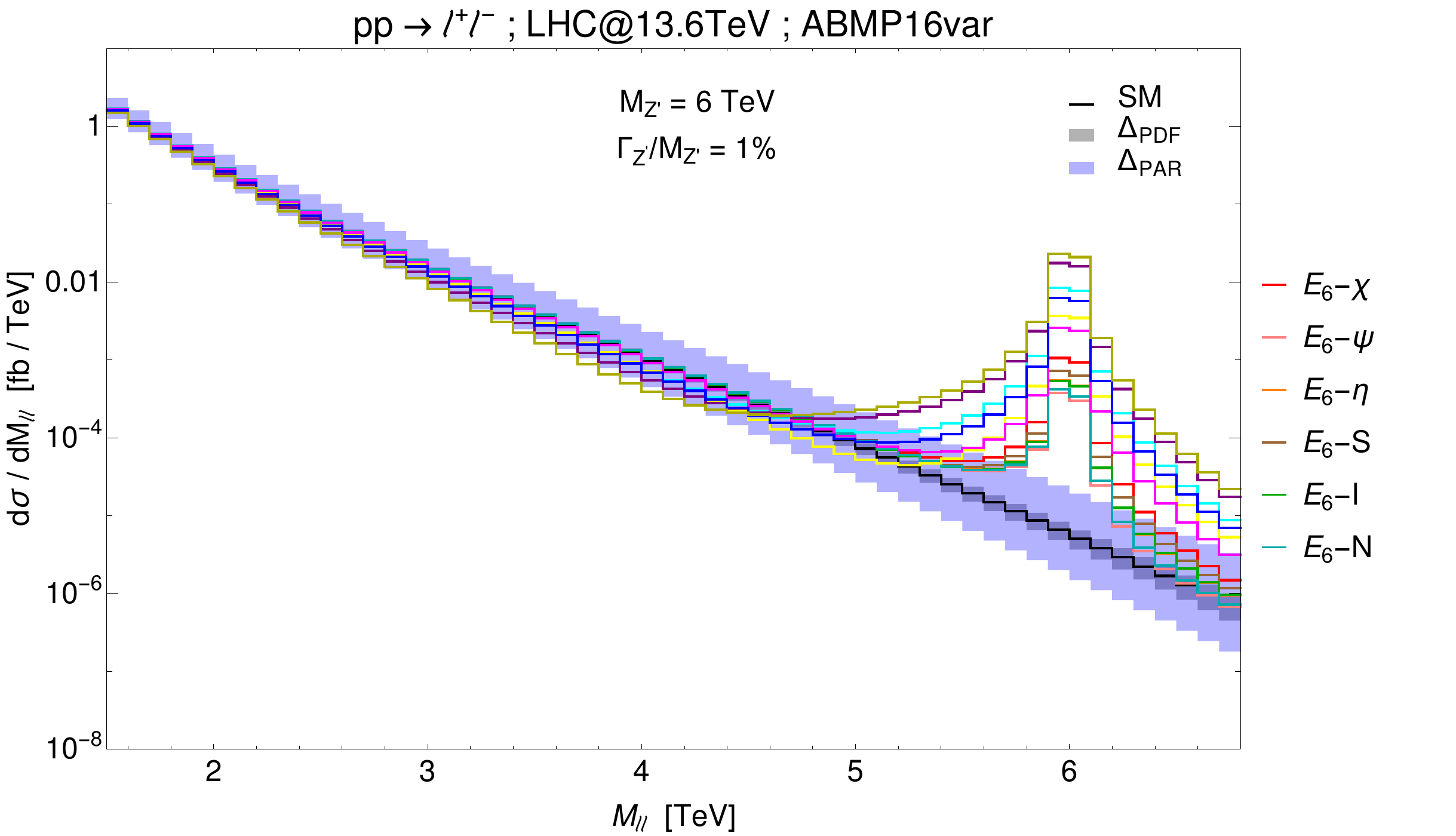}
\includegraphics[width=0.40\textwidth]{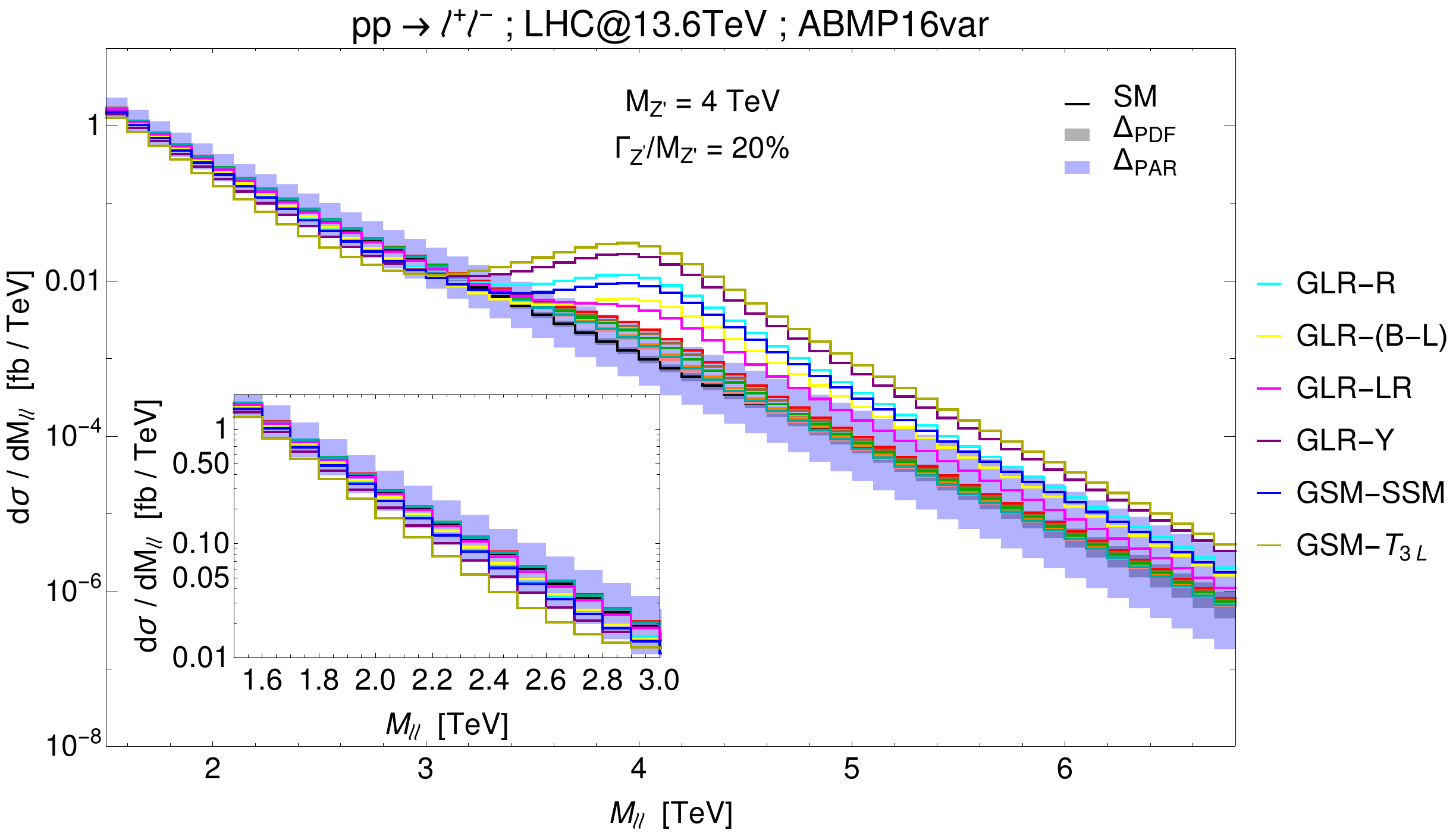}
\caption{Differential cross section in invariant mass of the dilepton final state at the LHC with $\sqrt{s} = 13.6$ TeV.
Coloured curves are obtained for a series of single $Z^\prime$ benchmark models, where we set (left) $M_{Z^\prime} = 6$ TeV, $\Gamma_{Z^\prime} / M_{Z^\prime} = $ 1\% and (right) $M_{Z^\prime} = 4$ TeV, $\Gamma_{Z^\prime} / M_{Z^\prime} = $ 20\%.}
\label{fig:bsm-invmass}
\end{center}
\end{figure}
\begin{figure}[t!]
\begin{center}
\includegraphics[width=0.40\textwidth]{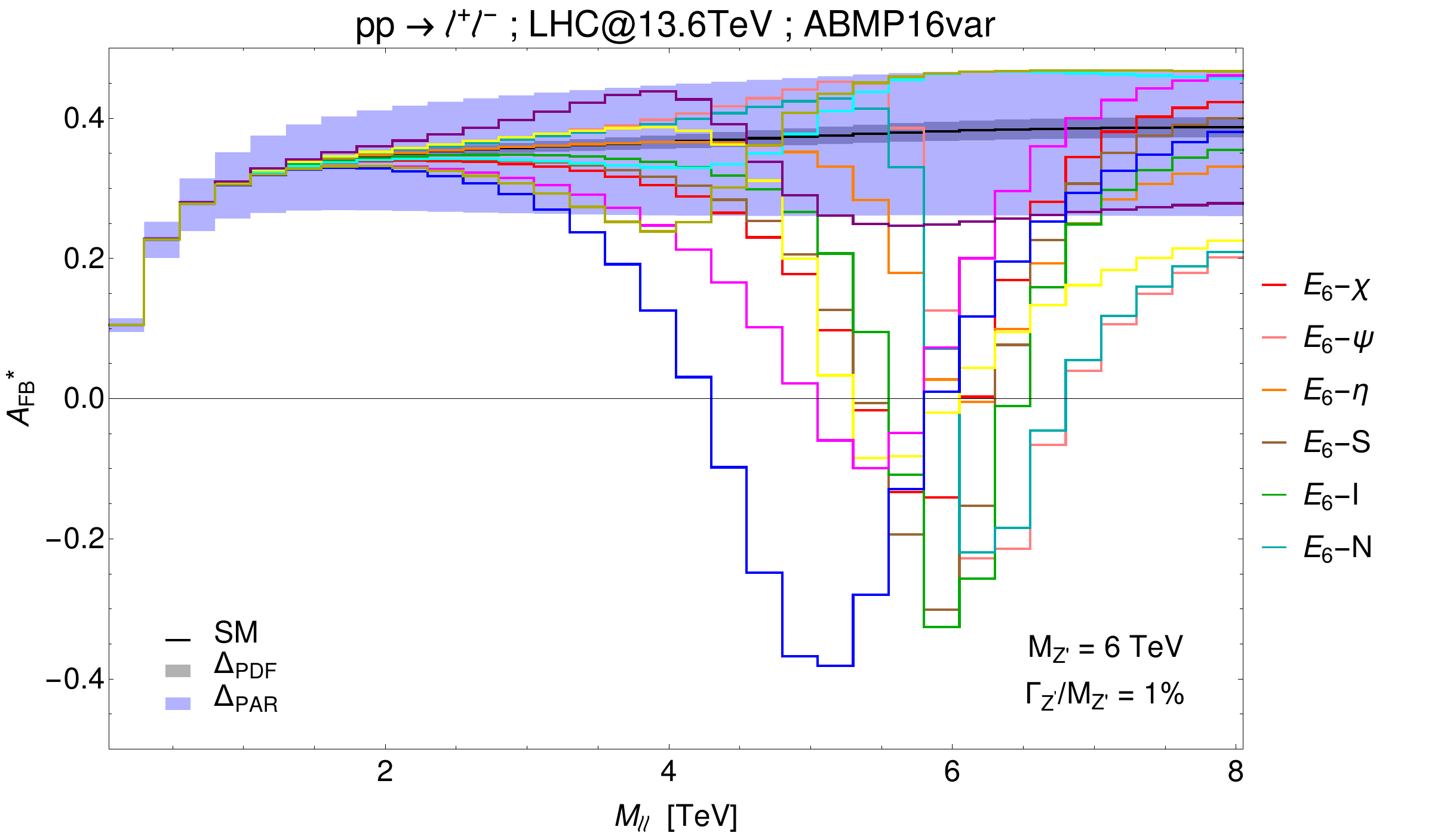}
\includegraphics[width=0.40\textwidth]{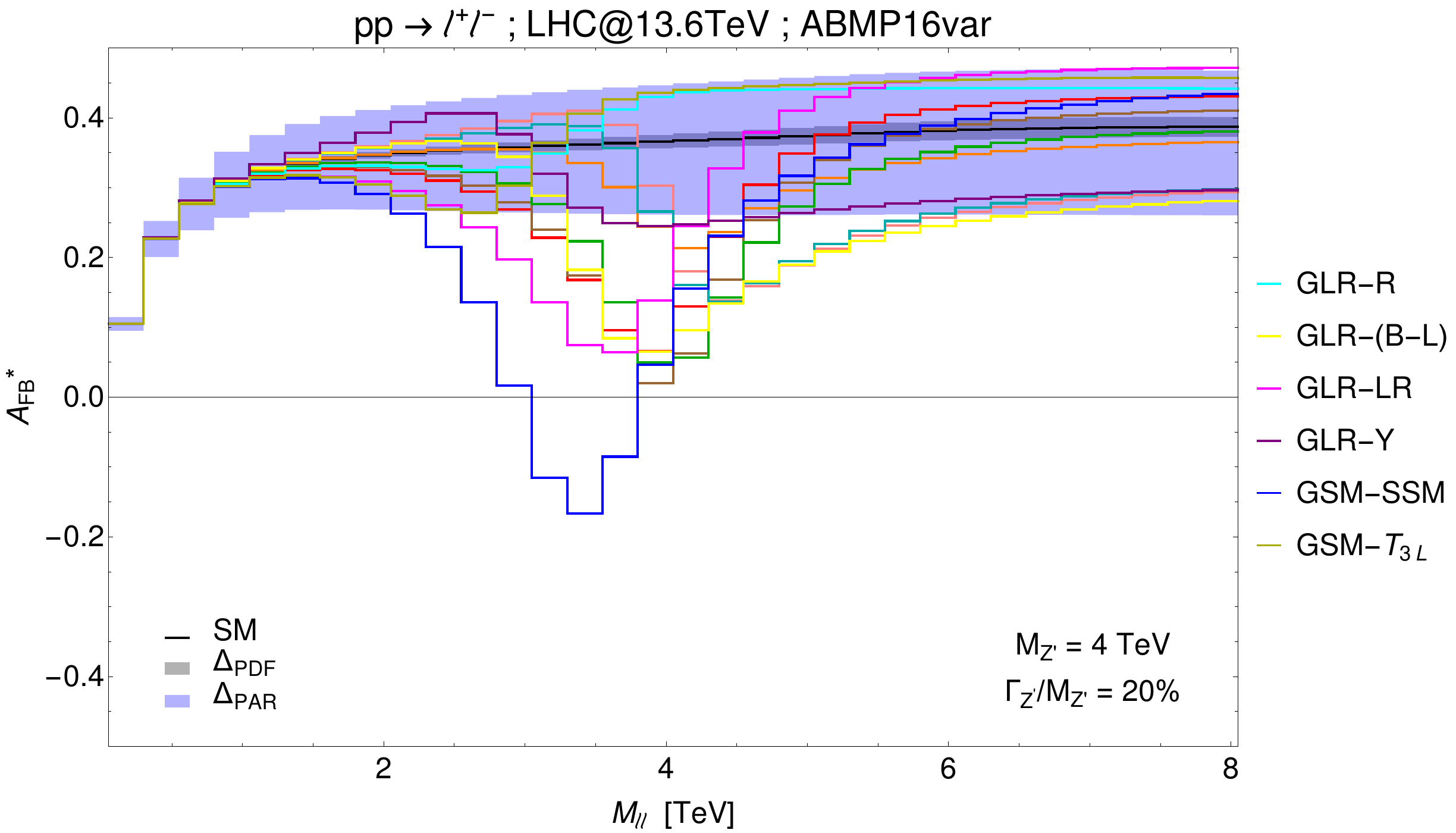}
\caption{Asymmetry 
$A_{\rm{FB}}^*$ in invariant mass of the dilepton final state 
at the LHC with $\sqrt{s} = 13.6$ TeV.
Coloured curves are obtained for a series of single $Z^\prime$ benchmark models, where we set (left) $M_{Z^\prime} = 6$ TeV, $\Gamma_{Z^\prime} / M_{Z^\prime} = $ 1\% and (right) $M_{Z^\prime} = 4$ TeV, $\Gamma_{Z^\prime} / M_{Z^\prime} = $ 20\%.
}
\label{fig:afb-bsm}
\end{center}
\end{figure}
\begin{figure}[t!]
\begin{center}
\includegraphics[width=0.40\textwidth]{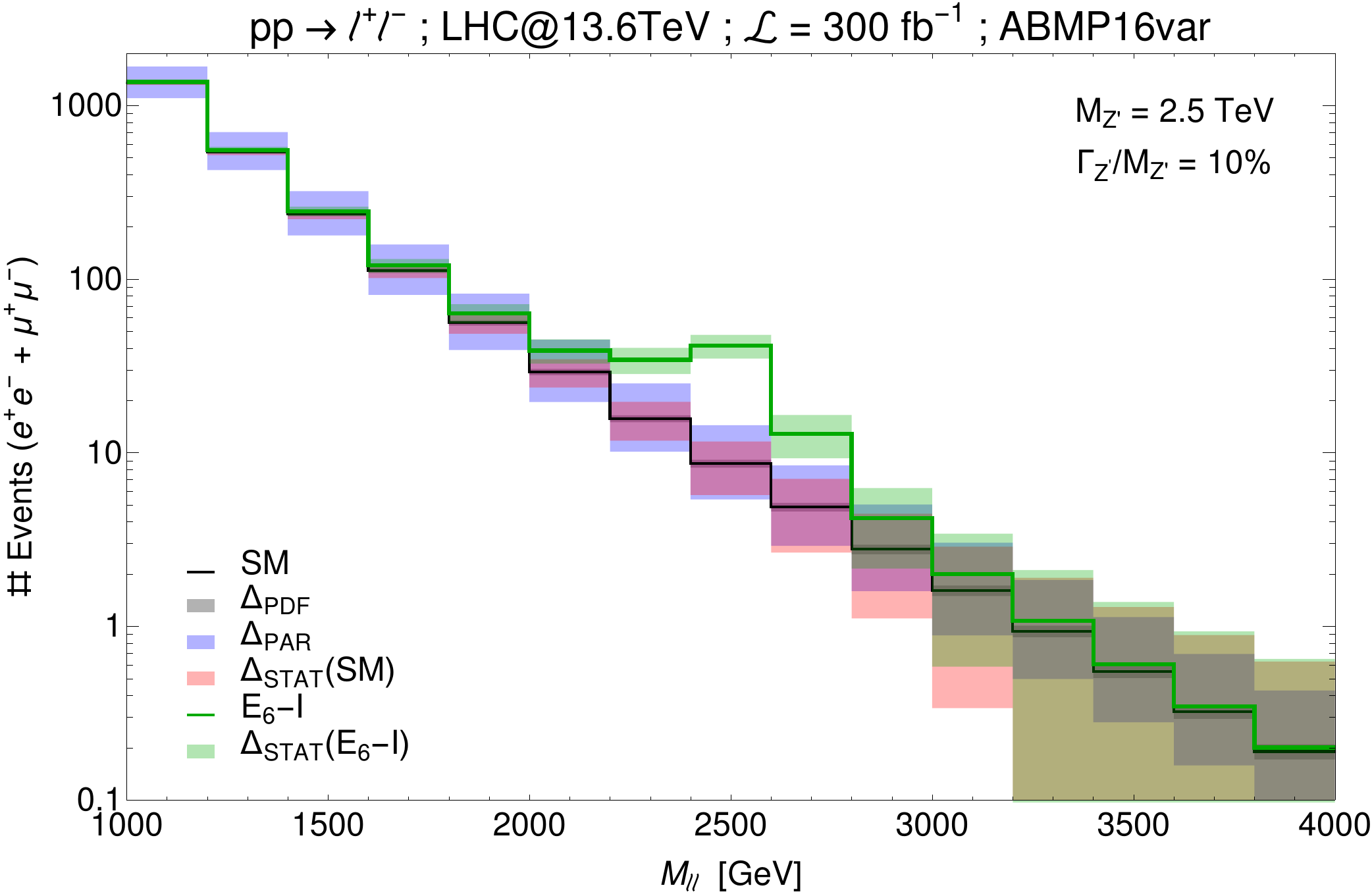}
\includegraphics[width=0.40\textwidth]{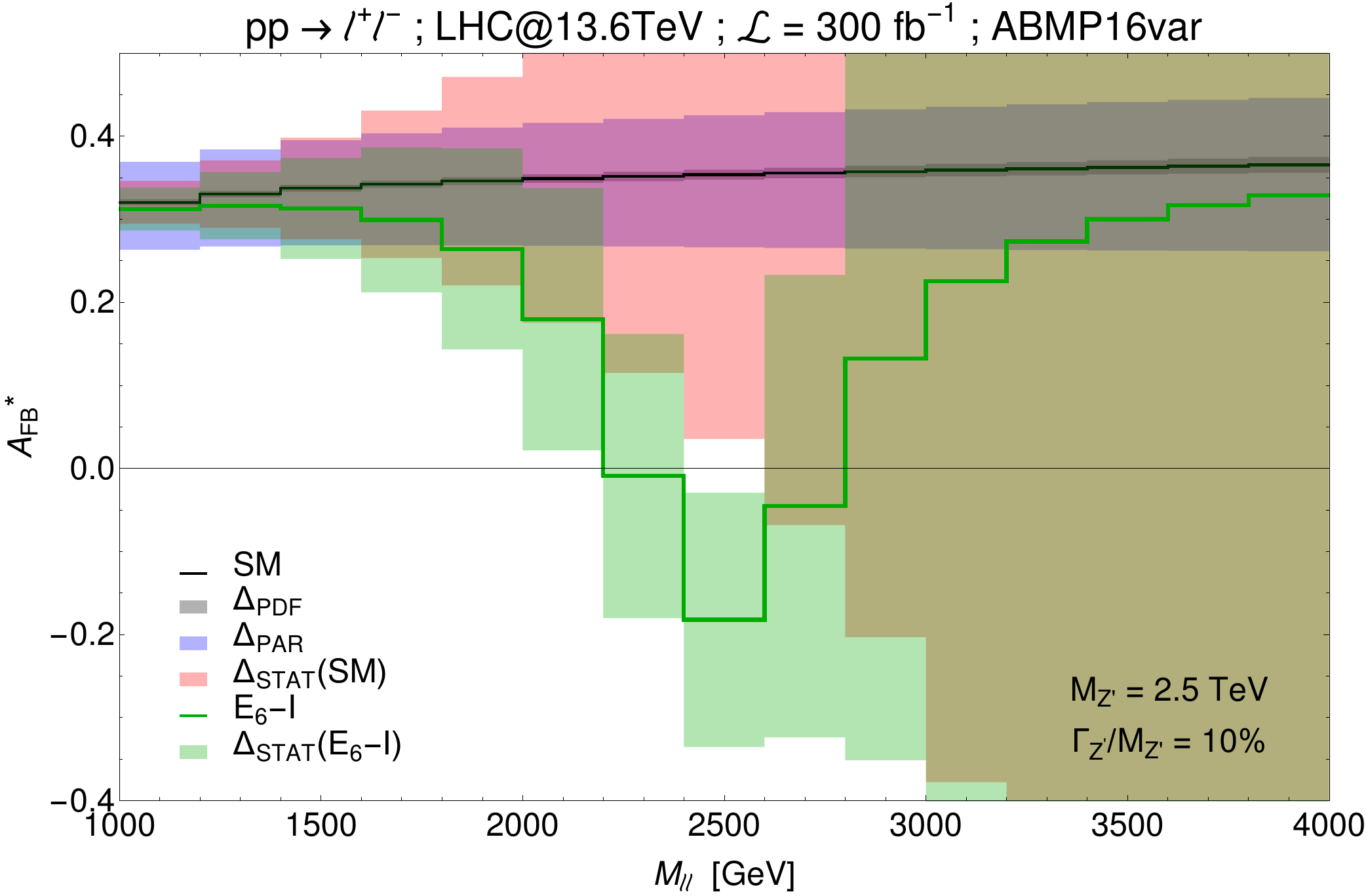}
\caption{Differential cross section (left) and $A_{\rm{FB}}^*$ (right) in invariant mass of the dilepton final state at the LHC with $\sqrt{s} = 13.6$ TeV for a $Z^\prime$ in the $E_6$-I model, where we set $M_{Z^\prime} = 2.5$ TeV and $\Gamma_{Z^\prime} / M_{Z^\prime} = $ 10\%.
The statistical uncertainty band corresponds to an integrated luminosity $\mathcal{L}$ = 300 fb$^{-1}$, combining the statistics of the di-electron and di-muon channels.}
\label{fig:bsm-benchmark}
\end{center}
\end{figure}

We will now investigate the systematic uncertainty from the high-$x$ quark density in the context of BSM searches,  as it is generally comparable to or larger than the ``standard'' PDF error.  The signal profile in the differential cross section observable for narrow (left) and wide (right) $Z^\prime$-bosons from a series of benchmark~\cite{Accomando:2010fz} is shown in Fig.~\ref{fig:bsm-invmass}.  Broad resonances would potentially suffer a strong reduction of sensitivity, while narrow models would be marginally affected by the additional source of uncertainty.  The negative interference contribution occurring in the low mass tail of the distribution appears more visibly in the former case. The resulting depletion of events can also lead to an early indication of the presence of a  BSM contribution~\cite{Accomando:2019ahs, Fiaschi:2021sin}, above and beyond SM uncertainties (for some $Z^\prime$ scenarios).\\
Fig.~\ref{fig:afb-bsm} shows the signal profiles in the $A_{\rm{FB}}^*$ observable of the selected benchmark models for the two configurations of narrow (left) and broad (right) resonances.
Despite the larger uncertainty band from the exponent variation method, the $A_{\rm{FB}}^*$ $Z^\prime$ signal shape remains well visible above the SM background predictions in both scenarios of narrow and broad resonances.
The property of the $A_{\rm{FB}}^*$ of being to some extent unaffected by variations of the resonance width is a crucial feature that makes this observable a suitable discriminant in BSM searches~\cite{Accomando:2015cfa}.\\
A statistical analysis over a selected $Z^\prime$ benchmark model is also performed to assess the sensitivity of the LHC to wide $Z^\prime$ resonances. The $E_6$-I model~\cite{Hewett:1988xc} is considered, where the $Z^\prime$ mass is set to 2.5 TeV and its width to 10\% of its mass.  The number of events (left) and the $A_{\rm{FB}}^*$ (right) with their statistical error in comparison with the systematic uncertainties from PDFs and from PDF parameterisation are shown in Fig.~\ref{fig:bsm-benchmark}.
Next-to-Next-to-Leading Order (NNLO) QCD corrections through a $K$-factor computed with \texttt{DYTurbo}~\cite{Camarda:2019zyx} are included, in order to obtain a more realistic estimation of the statistics of such a signal. The Electro-Weak (EW) parameters are fixed to the $G_\mu$ scheme at LO. The CMS experimental acceptances and efficiencies of the di-electron and di-muon channels~\cite{CMS:2021ctt} are included and the statistics of the two final states corresponding to an integrated luminosity of 300 fb$^{-1}$ are then combined.\\
A significance of 4.4$\sigma$ and 4.3$\sigma$ could be reached for the bump search and for the $A_{\rm{FB}}^*$ observable, respectively, when considering only the statistical uncertainty.
On the other hand,  the significance is reduced to 2.9$\sigma$ and 2.3$\sigma$ for the cross section and $A_{\rm{FB}}^*$ observable, respectively, once systematic uncertainties are included (linearly combining the two PDF errors while summing in quadrature the statistical uncertainties).
Therefore, in this context the significance from the $A_{\rm{FB}}^*$ still remains comparable to the cross section and, through the combination of the two, an earlier discovery can be achieved. 

\section{Conclusion}
The theoretical systematic uncertainties in the multi-TeV mass region associated with the high-$x$ quark density have been investigated. A method, to take into account the physical effects of propagating the low-scale non-perturbative parameterisation uncertainty in the falling-off $x \to 1$ quark density to the region of the very high mass tails in dilepton distributions has been developed  We have examined the implications of  this method both on SM predictions for  high-mass dilepton observables and  on BSM signals in a variety of narrow-$Z^\prime$ and wide-$Z^\prime$ models. We have further carried out a statistical analysis  of LHC Run 3 sensitivity to wide $Z^\prime$ resonances,  for integrated luminosity $\mathcal{L}$ = 300 fb$^{-1}$.

\end{document}